\documentclass[12pt]{amsart}
\usepackage{graphicx}
\usepackage{booktabs}
\usepackage{amsmath,amssymb,amsthm,graphicx,url}
\usepackage{fullpage}
\usepackage{listings,color}

\usepackage{times}
\usepackage{pdflscape}

\definecolor{dkgreen}{rgb}{0,0.6,0}
\definecolor{gray}{rgb}{0.5,0.5,0.5}
\definecolor{mauve}{rgb}{0.58,0,0.82}

\title{Introduction to the declination function for gerrymanders} 

\author{Gregory S. Warrington}
\address{Department of Mathematics \& Statistics, University of Vermont, 16 Colchester Ave., Burlington, VT 05401, USA}
\email{gregory.warrington@uvm.edu.}

\begin{document} 

\maketitle 


\begin{abstract}
  The \emph{declination} is introduced in~\cite{declination} as a new
  quantitative method for identifying possible partisan gerrymanders
  by analyzing vote distributions. In this expository note we explain
  and motivate the definition of the declination. We end by computing
  its values on several recent elections.
\end{abstract}


There are two main methods used for mathematically identifying
partisan gerrymanders. The first is to define functions that identify
oddly shaped districts under the assumption that unusual shapes are
likely due to gerrymandering. The second is to consider how votes
between the two parties are distributed among the districts. The
\emph{declination}, which we introduce in~\cite{declination} takes
this second approach. (Note that simulated district plans are
frequently used in conjunction with either or both of these methods.)
The purpose of this exposition is to provide an approachable
introduction to the declination, a method for measuring the degree of
a gerrymander. The declination treats asymmetry in the vote
distribution as indicative of gerrymandering. We refer the reader
to~\cite{declination} for references to some of the significant other
mathematical work on gerrymandering as well as a more comprehensive
analysis of elections using the declination.

\section{Brief Introduction}

We first briefly introduce the declination measure and mention some of
its strengths and weaknesses. The definition is motivated, and terms
are explained more fully, in the next section.

Plot the democratic vote fractions in each electoral district in increasing
order. Place three points on the diagram:
\begin{itemize}
  \item A point $F$ at the center of mass of the points corresponding to
    republican districts. The $y$-value (i.e., democratic vote
    fraction) of this point is the average of the $y$-values for these
    districts. The $x$-value is centered horizontally on the
    republican districts.
  \item A similar point $H$ corresponding to the democratic districts.
  \item A point $G$ whose $y$-value is at one-half and which
    horizontally resides at the transition between the republican
    districts and the democratic districts.
\end{itemize}
Draw the line segments $\overline{FG}$ and $\overline{GH}$. Compute
the angle between them (in radians). Then multiply by $2/\pi \approx
0.64$ so that the resulting value is between $-1$ and $1$. This is the
declination, $\delta$. Positive values indicate asymmetry that favors
Republicans while negative values indicate asymmetry that favors
Democrats. The name declination is in analogue to the angle between
true north and magnetic north.

\begin{figure}
  \centering
  \includegraphics[width=.8\linewidth]{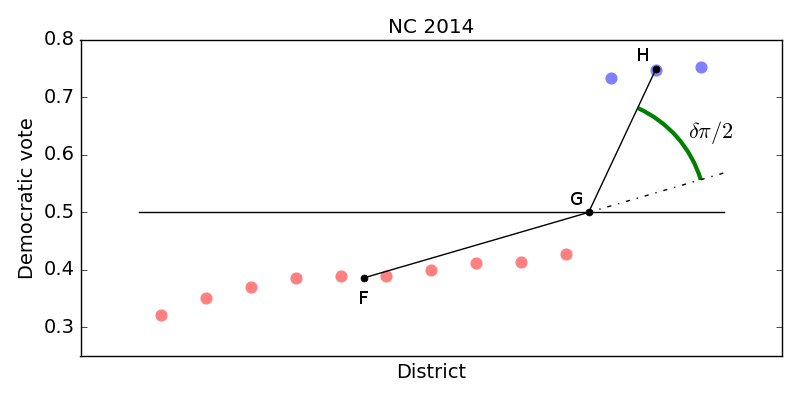}
  \caption{Example illustrating the three points $F$, $G$ and $H$
    arising in the definition of the declination. Data is from the
    2014 North Carolina election for the US House.}
\end{figure}

The following Python function computes the declination when given the
democratic vote fraction in each district as a list of numbers (each
between 0 and 1). The code to compute the declination and two variants
introduced in~\cite{declination}, along with R code to compute the
same, can be found at the author's website~\cite{homepage}.

\newpage

\begin{lstlisting}
import math
import numpy as np

def declination(vals):
    """ Compute the declination of an election.
    """
    Rwin = sorted(filter(lambda x: x <= 0.5, vals))
    Dwin = sorted(filter(lambda x: x > 0.5, vals))

    # Undefined if each party does not win at least one seat
    if len(Rwin) < 1 or len(Dwin) < 1:
        return False

    theta = np.arctan((1-2*np.mean(Rwin))*len(vals)/len(Rwin))
    gamma = np.arctan((2*np.mean(Dwin)-1)*len(vals)/len(Dwin))

    # Convert to range [-1,1]
    # A little extra precision just in case.
    return 2.0*(gamma-theta)/3.1415926535 
\end{lstlisting}

\vspace*{.2in}
\noindent
\textbf{Strengths of the declination}
\begin{enumerate}
\item Is a measure of partisan symmetry that does not assume any
  particular seats-votes proportionality.
\item Is a geometric angle that can be easily visualized on top of a
  plot of the votes among the various districts.
\item When scaled by half the number of districts, corresponds to the
  number of seats won by one side that are allocable to the asymmetry.
\item Can easily be used in conjunction with simulations to account
  for external sources of asymmetry such as geographic clustering.
\item Provably increases in absolute value in response to packing and cracking.
\item Continues to work even when one party is dominant statewide.
\item Is insensitive to incumbency gerrymandering (when done by both
  parties) as well as the degree of competitiveness of the election.
\end{enumerate}

\noindent
\textbf{Weaknesses of the declination}
\begin{enumerate}
\item Is not defined when one party sweeps all seats.
\item Is noisy when there are very few seats or when one party wins
  almost all of the seats (say, greater than $90\%$).
\end{enumerate}

\section{Motivation}

\begin{figure}
  \centering
  \includegraphics[width=.8\linewidth]{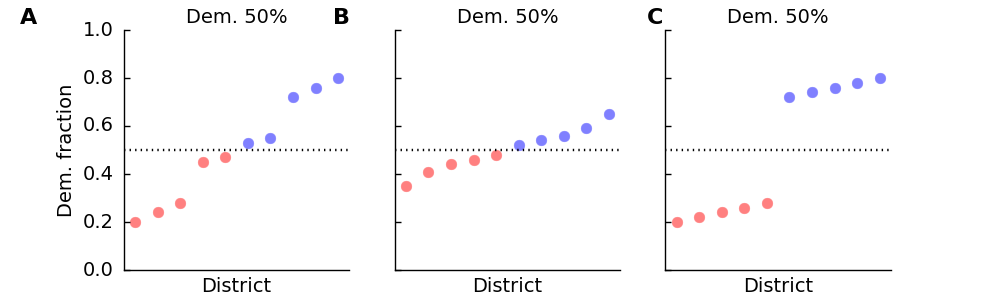}
  \caption{Typical, symmetric vote distributions for fair elections
    with equally popular parties.}
  \label{fig:intro1}
\end{figure}

To begin, suppose we have a state with ten electoral districts and
two major parties: the Democrats and the Republicans. The support of
each party will vary from district to district. Assuming the parties
are equally popular, there are probably a few districts in which the
Democrats are dominant; others in which the Republicans are dominant; and a few
where the races are likely to be competitive. If we write down the
fraction of Democrats voters in each district we get a sequence of ten
numbers, each between 0 and 1. A number close to zero means the Democrats
are a small minority in that district while a number close to 1
indicates they are overwhelming favorites. In Fig.~\ref{fig:intro1} we
have plotted the results for a few hypothetical elections for which
the parties are equally matched in the state overall. Each dot
corresponds to a single district. We have chosen to sort the districts
in increasing order of democratic vote. Doing so makes it easier to see what
is going on, but there is nothing magical about this ordering.

In Fig.~\ref{fig:intro1}A, we have the situation sketched in the
previous paragraph: the first three districts are dominated by the
Republicans, the last three by the Democrats, and the four in the
middle are dominated by neither party.  In Fig.~\ref{fig:intro1}B, we
have a similar scenario, except now there is much less variation from
district to district. The election results depicted are what you might
expect from sprinkling voters from either party down on the landscape
at random. By chance there will be some areas with a few more voters
of a given party, but overall the distribution of voters is relatively
homogeneous. In Fig.~\ref{fig:intro1}C there is a significant amount
of variation from district to district. For whatever reason, the
Democrats and Republicans are each clustered in five of the ten
districts.

\begin{figure}
  \centering
  \includegraphics[width=.5\linewidth]{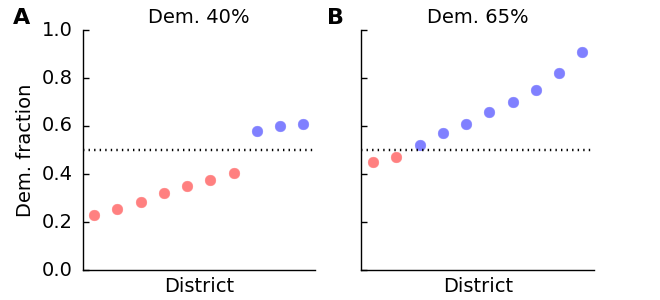}
  \caption{Plausible fair vote distributions when one party dominates.}
  \label{fig:intro2}
\end{figure}

In Figs.~\ref{fig:intro2}A and B we illustrate elections in which the
Republicans and the Democrats are the majority party,
respectively. The distributions still seem intuitively equitable. In
Fig.~\ref{fig:intro2}A, the Democrats have 40\% of the statewide vote
and win three of ten seats. While one might suppose that the Democrats
should win four seats, there isn't anything obviously unfair about how
the votes are distributed. The only unusual aspect is that none of the
districts are particularly competitive. The Democrats' votes are a
little more efficiently distributed than those of the Republicans in
that the Democrats win about 60\% of the vote in the districts they
win and the Republicans win about 68\% of the vote in the districts
they win. In Fig.~\ref{fig:intro2}B, the Democrats have 65\% of the
vote and win eight of the ten seats. In this case, there are some
competitive districts. In Fig.~\ref{fig:intro2}A we had two types of
districts --- Democrat-dominated and Republican-dominated. Here we
have a spectrum of districts, so it is harder to see if he
distribution is equitable. We can say, at the least, that there is a
reasonably continuous spectrum of districts ranging from narrow
Republican majorities to Democrat dominated.

\begin{figure}
  \centering
  \includegraphics[width=.8\linewidth]{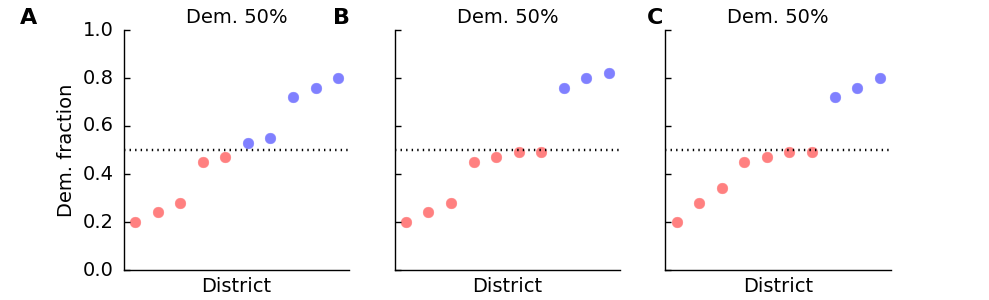}
  \caption{{\bf (A)} Repeat of election from Fig.~\ref{fig:intro1}A. {\bf (B)}
    Illustration of packing of the distribution from (A).  {\bf (C)}
    Illustration of cracking of the distribution from (A). }
  \label{fig:intro3}
\end{figure}

So far we have illustrated elections that are not, on their face,
obviously unfair to one of the parties. So how does one party get an
advantage? Partisan gerrymanders that advantage the Republicans at the
expense of the Democrats are created by ``packing and cracking'' the Democrat
voters. The most efficient way to win seats is through narrow
victories and, to the extent necessitated by overall support,
overwhelming defeats. When the Democrats win a district with an
overwhelming majority, it likely means there was another district that
could have been won by the Democrats had the voters been distributed more
evenly among the two districts. Likewise, two narrow losses by the
Democrats could likely have been one win and one loss had the votes been
distributed less evenly.

In Fig.~\ref{fig:intro3}B we have displayed what happens to the
election of Fig.~\ref{fig:intro1}A when extra Democrats are packed
into districts they were already going to win. The Democrats now only
win three districts; the ones they do win are won overwhelmingly. The
two additional districts the Republicans pick up are narrow wins, but
wins nonetheless. In terms of the plot of the vote-fractions, we see
that the dots for the districts the Democrats still win are further
away from the 50\% line while the dots for the two districts that
changed hands are just below 50\%. In Fig.~\ref{fig:intro3}C we show
an instance of cracking the election from Fig.~\ref{fig:intro1}A ---
votes are taken from districts the Democrats should have won and
distributed to other districts that they still have no hope of
winning. Once again, the Republican victories are, on average,
narrower than the Democrat victories.

\begin{figure}
  \centering
  \includegraphics[width=.5\linewidth]{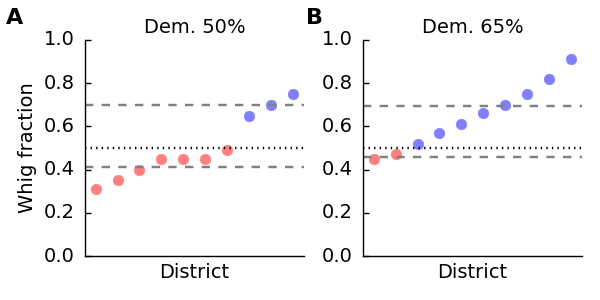}
  \caption{Distributions illustrating the limitations of comparing
    average margins of victory between the two parties.}
  \label{fig:intro4}
\end{figure}

As an initial attempt to determine whether the distribution of votes
is fair, we could compare the average democratic vote in the districts
the Democrats win to the average republican vote in the districts the
Republicans win (this is closely related to the \emph{lopsided means}
test suggested by Wang~\cite{Wang}). This works pretty well when the
parties are evenly matched statewide. Suppose the statewide average is
50\%. If the Democrats win less than half of the districts, their
average vote in those districts is necessarily higher than the average
vote for the Republicans in the districts the Republicans win. For
example, in Fig.~\ref{fig:intro4}A, the Democrats win only three
districts and do so with an average winning margin in these three
districts of 20\%. In contrast, the Republicans are able to win the
complementary seven districts with an average winning margin of about
8\%.

But simply comparing the two averages doesn't work as well when the
parties are not as closely matched in the state as a whole. Consider
Fig.~\ref{fig:intro2}B (repeated as Fig.~\ref{fig:intro4}B). The
Democrats have 65\% of the overall vote. It's not surprising at all
that the Republicans only win two seats. There has to be a fair amount
of geographic heterogeneity for there to be even two districts in
which the Republicans are the majority. Likewise, it's not surprising
that the average Democrat vote in the districts they win is far above
50\% while the Republican vote is barely above 50\% in the two
districts they do win. By the logic of the above paragraph, a simple
comparison of average winning margins would indicate that the
Democrats are grossly disadvantaged by the district plan since their
average winning margin is so much greater. But this is misleading. If
anything, they've won more districts then we might think they should
(65\% of the vote but 80\% of the seats).

\begin{figure}
  \centering
  \includegraphics[width=.8\linewidth]{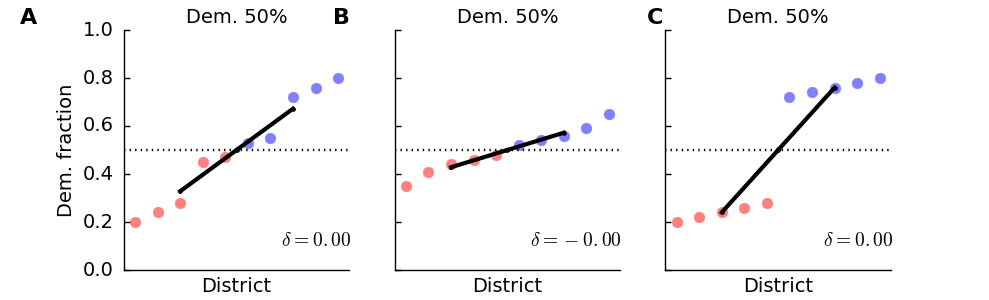}
  \caption{Fig.~\ref{fig:intro1} repeated with declination. As
    expected, the declination is 0 for each case since there is no
    partisan asymmetry.}
  \label{fig:intro1-dec}
\end{figure}

\begin{figure}
  \centering
  \includegraphics[width=.5\linewidth]{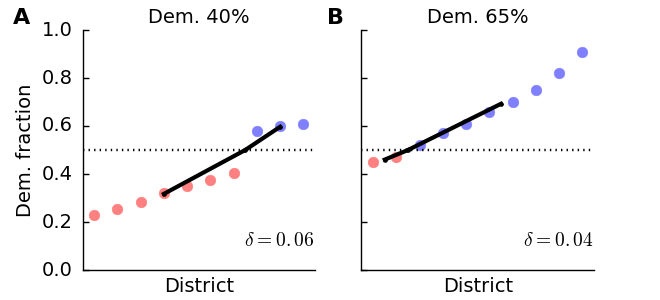}
  \caption{Fig.~\ref{fig:intro2} repeated with declination. The
    declination values are no longer 0, but they are safely below are
    (very arbitrary) threshold of 0.3.}
  \label{fig:intro2-dec}
\end{figure}

The declination addresses this by incorporating the fraction of seats
won into the comparisons of the averages. If a party only wins one or
two seats, we'd expect these wins to be relatively narrow. If the
party wins a lot of seats, we'd expect the average margin to be
relatively high. In light of this, the declination computes the ratio
of ``average winning margin'' to ``fraction of seats won'' for each
party. There's no particular assumption about an appropriate ratio;
the ratio will depend on how the supporters of each party are
distributed geographically. If the populations are distributed
relatively evenly across the state (lots of mixing of members of the
two parties), the ratio will be quite low. If there is little mixing,
the ratios are likely to be higher. Regardless, the underlying
assumption of the declination is that the ratios of the two parties
should be comparable. If not, then the parties are being treated
differently and one is getting an advantage from how the votes are
distributed.

\begin{figure}
  \centering
  \includegraphics[width=.8\linewidth]{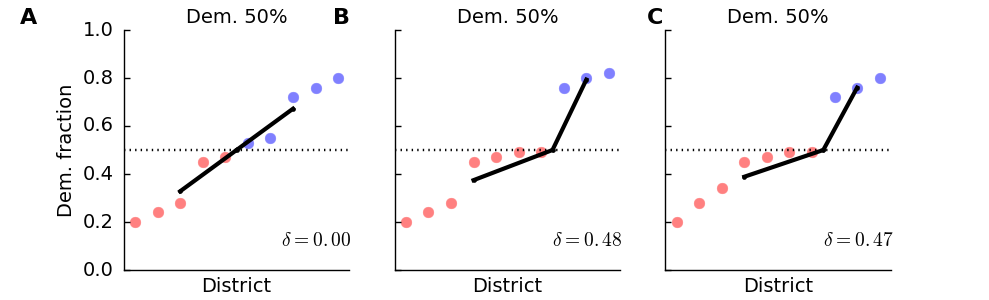}
  \caption{Fig.~\ref{fig:intro3} repeated with declination. Using our heuristic of
    multiplying by 5 (half the number of districts), we see that the
    declination estimates that in the second and third figures,
    approximately 2.4 and 2.35 seats have been turned republican by
    gerrymandering, respectively.}
  \label{fig:intro3-dec}
\end{figure}

For each party we thus have a right triangle: Its base is proportional
to (i.e., half of) the fraction of districts won and its height is
equal to the average margin of victory in the districts that are
won. This triangle has a hypotenuse of a given slope. \emph{The
  declination compares these slopes by computing the angle between
  lines of those slopes.} We refer the reader to~\cite{declination}
for an trigonometric expression of the definition. For now
it suffices to know that the declination ranges between $-1$ and $1$
with larger (absolute) values more indicative of partisan asymmetry.

As is suggested from the data in~\cite{declination}, \emph{for an
  election with 10 districts},
\begin{itemize}
  \item values above around $0.3$ could be considered indicative of
    likely gerrymandering (barring inherent geographic advantages) and
  \item multiplying the value of the declination by 5 estimates the
    number of seats switched due to gerrymandering. 
\end{itemize}

\begin{figure}
  \centering
  \includegraphics[width=.5\linewidth]{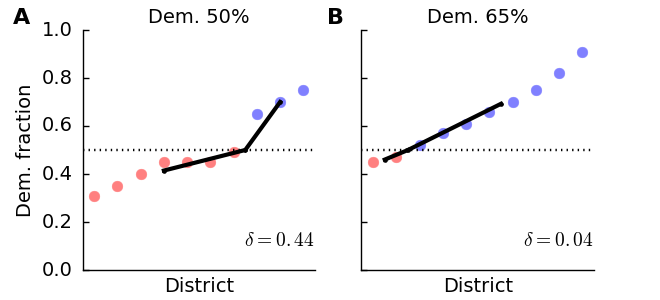}
  \caption{Fig.~\ref{fig:intro4} repeated with declination. The
    declination correctly identifies (B) as \emph{not} a gerrymander,
    unlike the simple comparison of averages used in the 
    discussion of Fig.~\ref{fig:intro4}B.}
  \label{fig:intro4-dec}
\end{figure}

Figs.~\ref{fig:intro1-dec}--\ref{fig:intro4-dec} repeat
Figs.~\ref{fig:intro1}--\ref{fig:intro4} but with the declination
shown (written as $\delta$). For each election, we have plotted the
hypotenuses of the two relevant triangles. The larger the difference
between the slopes, the larger (in absolute value) is the value of the
declination and the more unfair is the election.

\begin{figure}
  \centering
  \includegraphics[width=.8\linewidth]{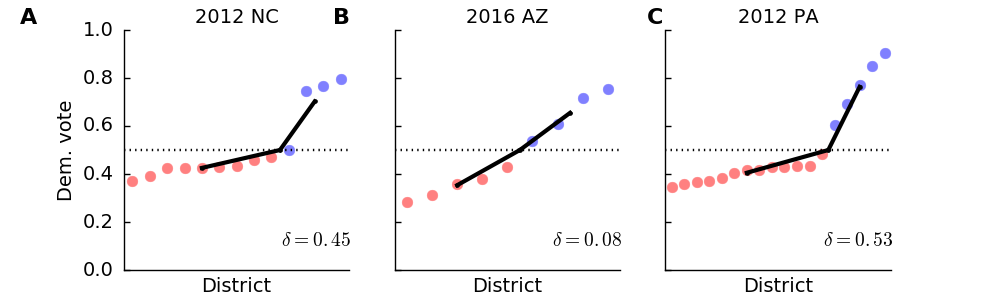}
  \caption{The declination for three recent elections.}
  \label{fig:intro5_dec}
\end{figure}

In Fig.~\ref{fig:intro5_dec} we see that the 2012 district plans for
North Carolina and Pennsylvania are advantageous to the Republicans
while the Arizona plan looks relatively neutral. (As described
in~\cite{declination}, we impute the vote fractions for uncontested
races.)  The heuristic for identifying the number of seats that have
changed parties is to multiply the declination by half the number of
districts. (See~\cite{seats} for a more sophisticated analysis of how
the declination relates to the number of seats switched.) For North
Carolina, this leads to an estimate of 2.9 seats and for Pennsylvania,
to an estimate of 4.8 seats.

As stated initially, the aim of this note is to give some intuition for
the motivation for the definition of the declination from a slightly
different perspective from that provided in our research
articles~\cite{declination,seats}. Our article~\cite{declination} in
particular provides more in-depth analysis including
\begin{itemize}
\item a theorem formally relating the declination with ``packing and cracking'',
\item a discussion of a variant of the declination that is more
  appropriate when one wishes to compare the declination for elections
  with different numbers of districts,
\item many more examples, and
\item a comparison to other quantitative measures of gerrymandering
  such as the efficiency gap~\cite{McGhee,M-S} and the mean-median
  difference~\cite{Wang,McDonaldBest}.
\item a discussion of confounding factors such as possible
  ``self-packing'' of Democrats and the Voting Rights Act of 1965.
\end{itemize}
Of course, to be useful as part of a manageable standard for
identifying candidate gerrymanders (beyond the scope of both this note
and~\cite{declination}), further validation analysis of the
declination must be done.

\bibliography{gerrymandering}

\begin{thebibliography}{{War}17b}

\bibitem[BW17]{seats}
J.~S. {Buzas} and G.~S. {Warrington}.
\newblock {Gerrymandering and the net number of US House seats won due to
  vote-distribution asymmetries}.
\newblock {\em ArXiv e-prints}, July 2017.
\newblock https://arxiv.org/abs/1707.08681.

\bibitem[MB15]{McDonaldBest}
Michael~D. McDonald and Robin~E. Best.
\newblock Unfair partisan gerrymanders in politics and law: A diagnostic
  applied to six cases.
\newblock {\em Elect. Law J.}, 14(4):312--330, Dec 2015.

\bibitem[McG14]{McGhee}
Eric McGhee.
\newblock Measuring partisan bias in single-member district electoral systems.
\newblock {\em Legis. Stud. Q.}, 39:55--85, 2014.

\bibitem[MS15]{M-S}
Eric McGhee and Nicholas Stephanopoulos.
\newblock Partisan gerrymandering and the efficiency gap.
\newblock {\em 82 University of Chicago Law Review}, 831, 2015.
\newblock 70 pages. U of Chicago, Public Law working Paper No. 493. Available
  at SSRN: https://ssrn.com/abstract=2457468.

\bibitem[Wan16]{Wang}
Samuel S.-H. Wang.
\newblock Three tests for practical evaluation of partisan gerrymandering.
\newblock {\em Stanford Law Review}, 68:1263--1321, June 2016.

\bibitem[War17a]{homepage}
Gregory~S. Warrington.
\newblock Python and r code for computing declination.
\newblock 2017.
\newblock \url{http://www.cems.uvm.edu/~gswarrin/research/}.

\bibitem[{War}17b]{declination}
Gregory~S. {Warrington}.
\newblock {Quantifying gerrymandering using the vote distribution}.
\newblock {\em ArXiv e-prints}, May 2017.
\newblock https://arxiv.org/abs/1705.09393. To appear, \emph{Election Law
  Journal}.

\end{thebibliography}

\bibliographystyle{alpha}

\end{document}